\documentclass[twocolumn,twocolappendix]{aastex6}

\begin{document}

\newcommand{\kms}{\ensuremath{\mathrm{km}\,\mathrm{s}^{-1}}}
\newcommand{\galunits}{\ensuremath{\mathrm{km}\,\mathrm{s}^{-1}\,\mathrm{kpc}^{-1}}}
\newcommand{\galacc}{\ensuremath{\mathrm{km}^2\,\mathrm{s}^{-2}\,\mathrm{kpc}^{-1}}}
\newcommand{\MLsun}{\ensuremath{\mathrm{M}_{\sun}/\mathrm{L}_{\sun}}}
\newcommand{\Lsun}{\ensuremath{\mathrm{L}_{\sun}}}
\newcommand{\Msun}{\ensuremath{\mathrm{M}_{\sun}}}
\newcommand{\Ha}{\ensuremath{\mathrm{H}\alpha}}
\newcommand{\SFR}{\ensuremath{\mathit{SFR}}}
\newcommand{\aveSFR}{\ensuremath{\langle \mathit{SFR} \rangle}}
\newcommand{\sfrate}{\ensuremath{\mathrm{M}_{\sun}\,\mathrm{yr}^{-1}}}
\newcommand{\Aunits}{\ensuremath{\mathrm{M}_{\sun}\,\mathrm{km}^{-4}\,\mathrm{s}^{4}}}
\newcommand{\surfdens}{\ensuremath{\mathrm{M}_{\sun}\,\mathrm{pc}^{-2}}}
\newcommand{\voldens}{\ensuremath{\mathrm{M}_{\sun}\,\mathrm{pc}^{-3}}}
\newcommand{\gevcc}{\ensuremath{\mathrm{GeV}\,\mathrm{cm}^{-3}}}
\newcommand{\etal}{et al.}
\newcommand{\LCDM}{$\Lambda$CDM}
\newcommand{\ML}{\ensuremath{\Upsilon_*}}
\newcommand{\Mst}{\ensuremath{M_*}}
\newcommand{\Mg}{\ensuremath{M_g}}
\newcommand{\Mb}{\ensuremath{M_b}}


\title{The Star Forming Main Sequence of Dwarf Low Surface Brightness Galaxies}

\author{Stacy S. McGaugh\altaffilmark{1}}
\author{James M. Schombert\altaffilmark{2}}
\author{Federico Lelli\altaffilmark{3}}
\altaffiltext{1}{Department of Astronomy, Case Western Reserve University, Cleveland, OH 44106, USA}
\altaffiltext{2}{Department of Physics, University of Oregon, Eugene, OR 97403, USA}
\altaffiltext{3}{European Southern Observatory, Karl-Schwarzschild-Strasse 2, 85748, Garching bei MŸnchen, Germany}

\begin{abstract}
We explore the star forming properties of late type, low surface brightness (LSB) galaxies.
The star forming main sequence (\SFR-\Mst) of LSB dwarfs has a steep slope, indistinguishable from unity ($1.04 \pm 0.06$).
They form a distinct sequence from more massive spirals, which exhibit a shallower slope.
The break occurs around $\Mst \approx 10^{10}\;\Msun$, and can also be seen in the gas mass---stellar mass plane.
The global Kennicutt-Schmidt law (\SFR-\Mg) has a slope of $1.47 \pm 0.11$ 
{without the break seen in the main sequence. There is an ample supply of gas in}
LSB galaxies, {which} have gas depletion times well in excess of a Hubble time, {and} often tens of Hubble times.
Only $\sim 3\%$ of this cold gas need be in the form of molecular gas to sustain the observed star formation. 
In analogy with the faint, long-lived stars of the lower stellar main sequence, 
it may be appropriate to consider the main sequence of star forming galaxies to be defined by 
thriving dwarfs (with $\Mst < 10^{10}\;\Msun$) while massive spirals (with $\Mst > 10^{10}\;\Msun$) are 
weary giants that constitute more of a turn-off population.
\end{abstract}


\section{Introduction}
\label{sec:Intro}

Considerable recent work has focussed on the star forming main sequence of galaxies \citep{NoeskeSFMS,Noeskegas}.  
Indeed, a dizzying array of calibrations of the star forming main sequence can be found in the 
literature \citep[e.g.][]{Peng2010,Speagle2014,Jaskot,Kurczynski,CanoDiaz}.
The fitted slope of the relation varies from rather flat at low redshift \citep[$\sim 0.5$:][]{Speagle2014} 
to rather steep at all redshifts \citep[$\sim 0.9$:][]{Kurczynski}.

The true form of the star forming main sequence has consequences for the ages and evolutionary states of actively
star forming galaxies. A slope of unity would be consistent with galaxies forming early in the universe and
subsequently forming stars at a nearly constant specific rate.
In contrast, shallow slopes imply that low mass galaxies have only recently formed, as
high specific star formation rates overproduce the observed stellar mass in less than a Hubble time.
These are drastically different pictures.

The terminology ``main sequence'' makes a clear parallel between star forming galaxies and the main sequence of stars.
While one may question the degree to which this analogy is appropriate, we note here that it may be carried further. 
The main sequence of stars for old systems is defined by slowly evolving, low-mass stars that linger long after
rapidly evolving, high-mass stars have evolved off the main sequence. Here we suggest that the star forming main sequence of galaxies 
should be defined by gradually evolving galaxies with gas depletion times that exceed the age of the Universe. 
Such systems are provided by late type, low surface brightness (LSB) galaxies \citep{ImpeyBothun1997,BothunLSB}.
In contrast, more commonly studied bright galaxies may represent a population that is currently ``turning off'' the main sequence. 

Low surface brightness (LSB) galaxies are ubiquitous in deep surveys that probe to faint isophotal 
levels \citep{LSBcatalog,gasrich,MihosUDG,vDUDG}. 
They share many properties with objects selected in blind HI surveys \citep{alfalfa}, which seem to identify the same population
\citep{alfalfaVirgo,LeoPdiscovery,almostdark}. 
LSB galaxies are generally late morphological types (Sd, Sm, Irr) spanning a large range in size and mass \citep{MSB95}. 
As the antithesis of the bright galaxies that dominate magnitude limited samples, they frequently provide new insight into
basic questions in galaxy formation and evolution. Star formation in LSB galaxies is an obvious topic of 
interest \citep[e.g.][]{vdH93,Boissier08,Wyder09,Meurer09,Lee09}

We have made a detailed study of LSB galaxies in a series of papers \citep{SMI,SMII,SMIII,SMIV,SMV}.
Our sample extends to very low masses ($\Mst < 10^7\;\Msun$) and star formation rates ($\SFR < 10^{-4}\;\sfrate$),
well below the $\Mst \sim 10^9\;\Msun$ lower limits typical of studies based on magnitude limited samples
and reaching the regime of single O stars.  
The large dynamic range in stellar mass observed in LSB galaxies provides a new constraint on the slope
of the star forming main sequence.
We also consider the relation of star formation in LSB galaxies with gas mass and baryonic mass, 
and note that the relation between stellar and gas mass contains similar information to the star forming main sequence.

\section{Data}
\label{sec:data}

Our sample is drawn from the LSB galaxy catalog of \citet{LSBcatalog} and the dwarf LSB catalog of \citet{gasrich}.
The specific sample of galaxies considered here consists of the 56 LSB galaxies listed in Table \ref{sfrdata} for which a great deal of data are available. 
These include gas masses, optical luminosities, colors \citep{SMI} and \Ha\ luminosities \citep{SMI,SMII}. 
In some cases, near-IR luminosities are also available \citep{SMIV}. 
These narrowband and broadband photometric data provide a picture of star formation in a population of LSB galaxies 
covering a large range in mass, $5 \times 10^6 < \Mst < 7 \times 10^9 \;\Msun$.
The low end of this mass range overlaps with the stellar masses of the dwarf spheroidal satellites of the Local Group,
while the upper end approaches (though does not quite reach) that of $L^*$ spiral galaxies.

There is modest overlap with previous work: eight objects in common with \citet{hunter}; 
ten in common with \citet{Wyder09} and one object in common with \citet[][]{Boissier08}.
There is reasonable agreement between independent data sets where there is overlap \citep[see discussion in][]{SMII}.
Masses and star formation rates generally agree well once differences in method and calibration are accounted for.

\begin{table}[t!]
\begin{deluxetable}{lrcccc}
\tablewidth{0pt}
\tablecaption{LSBG Masses and Star Formation Rates\label{sfrdata}}
\tablehead{
\colhead{Galaxy}  & \colhead{D} & \colhead{$\log(\Mst)$} & \colhead{$\log(\Mg)$} & \colhead{$\log(\Mb)$} & \colhead{$\log(\SFR)$} \\
& (Mpc) & \multicolumn{3}{c}{$(\mathrm{M}_{\sun})$} & \colhead{($\mathrm{M}_{\sun}\;\mathrm{yr}^{-1}$)} 
 }
\decimals
\startdata
F615-1	&8.2	&6.76  &7.55 & 7.61  &-4.38 \\[-4pt]
F750-V1	&8.0	&6.77  &7.14 & 7.30  &-3.89 \\[-4pt]
F565-V1	&10.8	&6.94  &7.00 & 7.27  &-4.14 \\[-4pt]
F533-1	&12.8	&7.24  &8.11 & 8.16  &-3.30 \\[-4pt]
F608-1	&9.0	&7.30  &7.73 & 7.87  &-3.35 \\[-4pt]
DDO 154	&4.04	&7.43  &8.59 & 8.62  &-2.65 \\[-4pt]
F608-V1	&20.3	&7.61  &8.56 & 8.61  &-3.01 \\[-4pt]
D570-7	&15.6	&7.63  &7.93 & 8.11  &-3.14 \\[-4pt]
D568-2	&21.3	&7.66  &7.95 & 8.13  &-2.84 \\[-4pt]
D646-5	&18.3	&7.69  &8.44 & 8.51  &-2.45 \\[-4pt]
F512-1	&14.1	&7.74  &8.20 & 8.33  &-2.55 \\[-4pt]
D646-7	&29.4	&7.76  &8.07 & 8.24  &-1.67 \\[-4pt]
D646-9	&7.2	&7.76  &8.62 & 8.67  &-2.85 \\[-4pt]
D575-7	&18.1	&7.81  &8.43 & 8.52  &-2.18 \\[-4pt]
D631-7	&15.9	&7.85  &8.48 & 8.58  &-1.95 \\[-4pt]
D656-2	&7.8	&7.85  &8.54 & 8.62  &-2.55 \\[-4pt]
D572-5	&18.0	&7.88  &8.37 & 8.49  &-2.31 \\[-4pt]
D637-3	&35.2	&7.95  &8.80 & 8.86  &-2.03 \\[-4pt]
D646-11	&12.1	&7.97  &7.99 & 8.28  &-2.47 \\[-4pt]
F544-1	&28.5	&7.97  &9.08 & 9.11  &-2.62 \\[-4pt]
F415-3	&10.4	&8.04  &8.65 & 8.75  &-2.68 \\[-4pt]
F611-1	&25.5	&8.06  &8.51 & 8.65  &-2.37 \\[-4pt]
D572-2	&56.5	&8.07  &8.88 & 8.94  &-1.88 \\[-4pt]
D570-3	&23.7	&8.09  &7.93 & 8.32  &-2.57 \\[-4pt]
DDO 168	&5.2	&8.11  &8.78 & 8.87  &-2.01 \\[-4pt]
D500-3	&22.7	&8.13  &8.25 & 8.50  &-2.14 \\[-4pt]
D495-2	&33.4	&8.18  &8.39 & 8.60  &-1.86 \\[-4pt]
D723-4	&32.9	&8.28  &9.10 & 9.16  &-1.98 \\[-4pt]
D495-1	&34.9	&8.30  &8.20 & 8.55  &-2.82 \\[-4pt]
F473-V2	&44.7	&8.30  &8.95 & 9.04  &-2.61 \\[-4pt]
F565-V2	&55.1	&8.31  &8.99 & 9.07  &-2.26 \\[-4pt]
D575-2	&14.7	&8.39  &8.77 & 8.93  &-2.02 \\[-4pt]
F563-V1	&57.6	&8.42  &8.94 & 9.06  &-2.25 \\[-4pt]
F612-V3	&65.4	&8.43  &9.06 & 9.15  &-1.48 \\[-4pt]
F583-2	&25.4	&8.54  &8.99 & 9.12  &-1.85 \\[-4pt]
F750-2	&46.1	&8.55  &9.28 & 9.35  &-1.80 \\[-4pt]
D563-1	&61.6	&8.58  &8.82 & 9.02  &-1.88 \\[-4pt]
F614-V2	&51.3	&8.60  &9.06 & 9.19  &-1.38 \\[-4pt]
D723-5	&27.7	&8.71  &8.43 & 8.89  &-1.71 \\[-4pt]
F651-2	&27.5	&8.73  &9.00 & 9.19  &-1.71 \\[-4pt]
F563-1	&52.2	&8.86  &9.66 & 9.73  &-1.13 \\[-4pt]
F677-V2	&63.9	&8.95  &8.85 & 9.20  &-1.75 \\[-4pt]
F687-1	&47.5	&8.98  &9.32 & 9.48  &-1.83 \\[-4pt]
D723-9	&26.2	&9.08  &8.99 & 9.34  &-1.45 \\[-4pt]
D564-9	&46.0	&9.17  &9.56 & 9.71  &-0.61 \\[-4pt]
UGC 5005	&57.1	&9.17  &9.84 & 9.93  &-0.69 \\[-4pt]
F562-V1	&68.1	&9.19  &9.50 & 9.67  &-0.98 \\[-4pt]
F574-2	&84.8	&9.30  &9.29 & 9.60  &-1.25 \\[-4pt]
F568-V1	&92.3	&9.30  &9.70 & 9.85  &-0.97 \\[-4pt]
F577-V1	&113.	&9.39  &9.78 & 9.93  &-0.72 \\[-4pt]
F561-1	&69.8	&9.43  &9.26 & 9.65  &-0.85 \\[-4pt]
F568-1	&95.5	&9.47  &9.72 & 9.91  &-0.95 \\[-4pt]
F574-1	&100.	&9.52  &9.73 & 9.94  &-0.59 \\[-4pt]
UGC 128	&58.5	&9.73 &10.13 &10.27  &-0.58 \\[-4pt]
D774-1	&72.0	&9.77  &9.67 &10.02  &-1.41 \\[-4pt]
F579-V1	&90.5	&9.84  &9.49 &10.00  &-1.07 \\[-4pt]
\enddata
\end{deluxetable}
\end{table}

To place our results in context, we utilize data from SINGS \citep{K09} and THINGS \citep{leroy}. 
These provide a reference sample that includes bright star forming galaxies with $\Mst > 10^{10}\;\Msun$ as
well as some dwarfs overlapping the mass range of our LSB galaxies.  These particular comparison samples
are used because they provide the same type of data that allows us to compare properties directly (e.g., 
star formation rates are consistently estimated from extinction corrected \Ha\ measurements).
No claim to completeness of these samples is made. We merely wish to place our sample of LSB galaxies
in the context of well studied local star forming galaxies.

Table \ref{sfrdata} lists the name, adopted distance, stellar mass, gas mass,
baryonic mass ($M_b = \Mst + \Mg$), and star formation rate for each galaxy.
These quantities are computed as described below.
Distances are adopted as described in \citet{SMI} with the exception of DDO 154, for which
we adopt the TRGB distance of \citet{EDD09}.
Uncertainties in distance dominate over photometric uncertainties in most cases,
but vary in the same way on the quantities of interest so should not affect the slope
of fitted relations: $M \propto D^2$. 

\subsection{Stellar Masses}

We estimate stellar masses from the observed luminosity and color such that $\Mst = \ML^V L_V$
using the mass-to-light ratio given by \citet{portinari} for a Kroupa \citep{kroupa} initial mass function (IMF):
\begin{equation}
\log \ML^V = 1.29(B-V)-0.654.
\label{eq:portinari}
\end{equation}
This model is practically indistinguishable from that of \citet{Bell03} and is consistent with other models \citep{MS14}.
Stellar masses estimated by applying eq.\ \ref{eq:portinari} to the data in \citet{SMI} are given in Table \ref{sfrdata}.

\begin{figure}
\epsscale{1.0}
\plotone{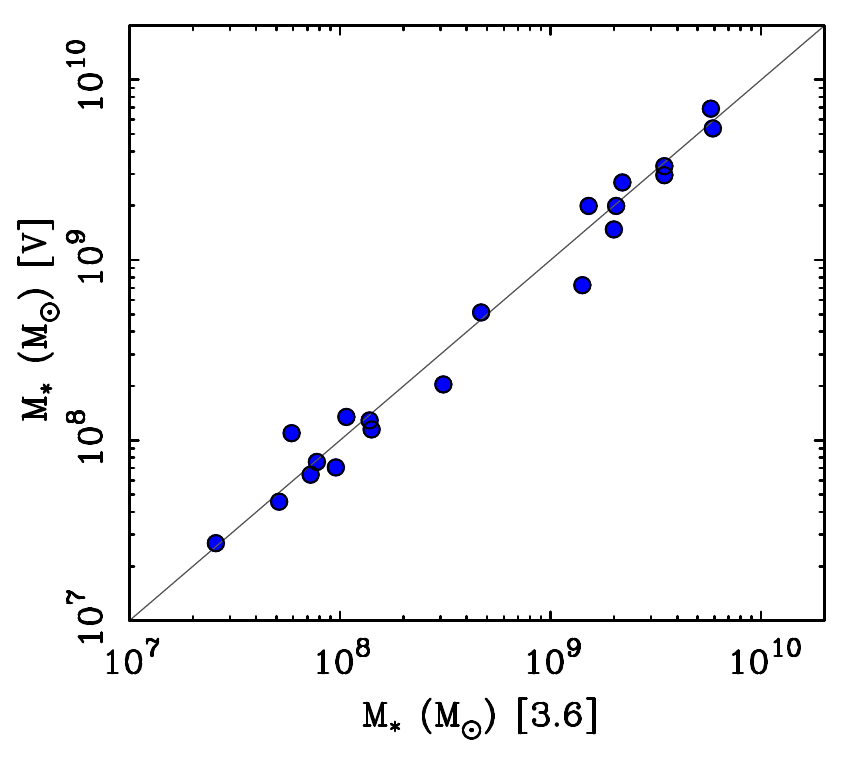} 
\caption{Stellar masses of LSB galaxies (points) with both $V$-band and 3.6$\mu$m data. 
The line is the line of unity.  Stellar mass estimated using optical $V$-band luminosities and $B-V$ colors (ordinate) 
compare well with those utilizing [3.6] Spitzer luminosities \citep[abscissa; see][]{MS14}, the scatter being $\sim 0.1$ dex.
\label{mstmst}}
\end{figure}

\citet{SMIII} constructed new population synthesis models while 
\citet{MS14} considered the applicability of the mass-to-light ratio estimators of published population synthesis models
to a broad range of photometric data. The upshot of this work is that the near-infrared bands (e.g., $K$ and [3.6]) provide the 
closest link to stellar mass, as expected. Due to the vagaries of  telescope scheduling and weather, only 20 of the
56 galaxies in Table \ref{sfrdata} with \Ha\ measurements \citep{SMI,SMII} also have Spitzer [3.6] images \citep{SMIV},
while all have $B$ and $V$ band observations enabling use of eq.\ \ref{eq:portinari}.

For the galaxies with all relevant data,
we compare the stellar mass estimated from eq.\ \ref{eq:portinari} to that from the [3.6] luminosity in Fig.\ \ref{mstmst}.
For the models of \citet{portinari}, \citet{MS14} find that the appropriate mass-to-light ratio at $3.6\mu$ is a constant
$\ML^{[3.6]} = 0.49\;\MLsun$, consistent with the $0.5\;\MLsun$ found by \citet{SMIII} and other population 
modeling \citep{eskew,meidt14,BeingWiseI}.  Fig.\ \ref{mstmst} shows that optical and near-IR stellar mass estimators agree reasonably
well. The data follow the line of equality, which in itself is a non-trivial accomplishment \citep[see][]{MS14}.
The rms scatter is 0.1 dex, which is arguably less than the uncertainty in the adopted IMF.

\subsection{Gas Masses}
\label{sec:Mg}

Gas masses are taken to be the HI masses \citep{eder,SMI} corrected for the cosmic abundance of helium
($\Mg = 1.33 M_{HI}$) and are given in Table \ref{sfrdata}. Molecular gas masses are unknown, and probably small.
Emission lines like CO have not been detected in the vast majority of LSB galaxies \citep{SBIM90,dBvdH98b,Cao17}.
There are detections in a couple of exceptional cases \citep{Das06,Das10}, but these
are giant LSB galaxies that are very different from the dwarfs discussed here \citep{LelliGaintLSB}.
 
\begin{figure*}
\epsscale{1.0}
\plotone{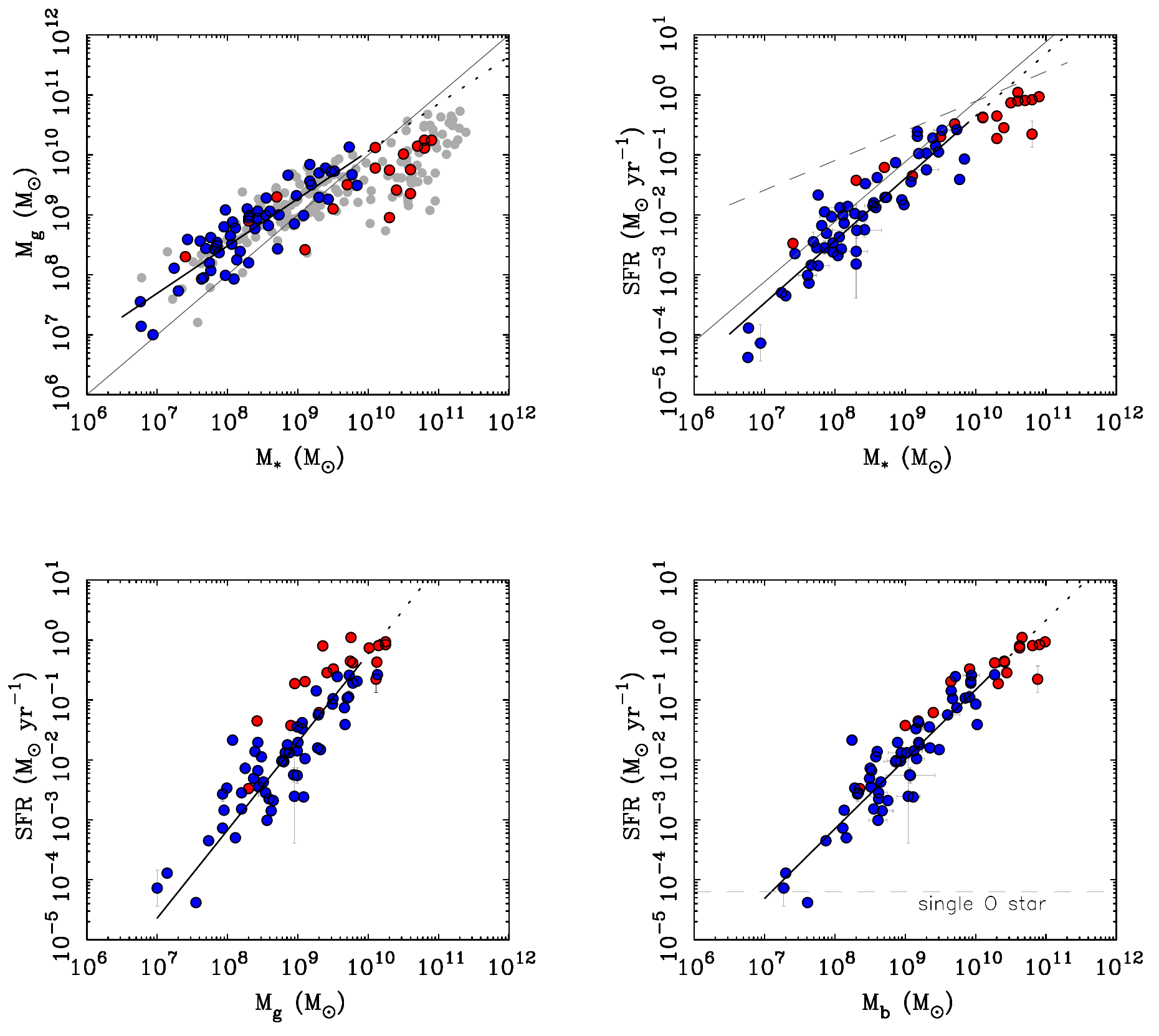}
\caption{Masses and star formation rates of local star forming galaxies: gas mass vs.\ stellar mass (top left), SFR vs.\ stellar
mass (top right: the star forming main sequence), SFR vs.\ gas mass (bottom left: the global Kennicutt-Schmidt relation), and
SFR vs.\ baryonic mass (bottom right). LSB galaxies for which we have measured \Ha\ fluxes are shown as blue points \citep{SMI}.
The dark solid lines are fits (Table \ref{fitresults}) to the LSB galaxy data (Table \ref{sfrdata}); extrapolated extensions of these fits are shown as dotted lines.
For comparison, SINGS galaxies \citep{K09} are shown as red points; these are not included in the fits.  
The range of properties of known disk galaxies is illustrated by the SPARC sample \citep[][gray points in top left panel]{SPARC}.
The light solid line in the top left panel is the line of equal mass; in the top right panel it is the line of constant SFR for 13 Gyr. 
Also shown at the top right is the main sequence fit of \citet[dashed line]{Speagle2014}. 
In the lower right panel we note the approximate threshold where the observed \Ha\ flux may be provided by a single O star.
Note the change in behavior at $\Mst \approx 10^{10}\;\Msun$ that separates the regime of thriving dwarfs and weary giants in the top panels.
\label{SFMS}}
\end{figure*}

To check the importance of molecular gas to the mass budget, we estimate the amount of molecular
gas mass based on the observed star formation rate (\S \ref{sec:SFR}) using eq.~1 of \citet{MS15}. 
This is obtained by assuming that LSB galaxies form stars with the same efficiency as other disk galaxies \citep{leroy}.
In effect, we simply ask how much molecular gas is required to sustain the observed star formation.
The answer is not much. On average, the mass in atomic gas exceeds that in molecular gas by a factor of 33.
This is a negligible contribution to the mass budget and we do not consider it further for the LSB sample.
For the comparison SINGS sample, we include the molecular gas component when it has been measured \citep{leroy}.

\subsection{Star Formation Rates}
\label{sec:SFR}

In this work we employ the \Ha\ line to measure star formation rates.
Accurate \Ha\ flux measurements are available for many LSB galaxies \citep{SMI,SMII}.
These data are in good agreement with independent measurements of the same galaxies where they have been made \citep[e.g,][]{hunter}.
Moreover, they are very deep, often detecting HII regions powered by single O stars.
In contrast, UV and far-IR surveys often fail to detect these dim galaxies, making \Ha\ the only viable probe of star formation.
We adopt the \Ha\ calibration of \citet{KE12}:
\begin{equation}
\log(\SFR) = \log[L(\Ha)] -41.27 
\label{eq:KEsfr}
\end{equation}
where the star formation rate is in \sfrate\ and the \Ha\ luminosity is in erg s$^{-1}$.

\subsubsection{Corrections for Extinction and [N II] Contamination}

We correct the observed \Ha\ fluxes for distance, extinction, and contamination by the lines of [N II] in the passband of the \Ha\ filter.
LSB galaxies generally have low dust content \citep{M94,Wyder09} and weak [N II] emission \citep{M94,dBvdH98a,KdN04}, so these corrections are much
smaller than in more commonly studied bright galaxies. Consequently, we expect \Ha\ to be a better tracer of star formation in
LSB galaxies than in dustier, higher metallicity spiral galaxies, at least insofar as the necessary corrections are smaller.

We use the spectroscopic observations of HII regions in LSB galaxies \citep{M94,dBvdH98a,KdN04} to estimate corrections
for extinction and [N II] contamination. The extinction is estimated from the Case B Balmer decrement ($\Ha/\mathrm{H}\beta = 2.8$).
The [N II]/\Ha\ ratio is measured directly in data with sufficient spectral resolution to cleanly separate them. 

While there is a good amount of spectroscopic data available, we do not have spectroscopy for all the HII regions in all of LSB galaxies for 
which we have \Ha\ luminosities. This makes it impossible to make a correction for each individual HII region. Moreover, the uncertainties on
these measurements for any one HII region can be large. We therefore make a {single} correction for the entire population.

{To obtain a mean correction for internal extinction, we start by noting that}
there is a clear trend of extinction with metallicity as traced by oxygen abundance \citep[Fig. 6 of][]{M94}.
{This goes} in the expected sense {that higher metallicity systems are dustier.}
There is very little extinction in the lowest metallicity HII regions, while there is clearly some in HII regions that are merely low abundance.
Splitting the data into bins above and below $12+\log(\mathrm{O/H}) = 8.1$, we obtain median extinction corrections of $A_{\Ha} = 0.23$ (lower Z) and
0.42 (higher Z) magnitudes.

We use the same metallicity bins to estimate a correction for [N II]. For LSB galaxies of higher metallicity, the mean $\mathrm{[N~II]}/\Ha = 0.12$.
In the lower metallicity bin, the median ratio is 0.06. We note that many of the lowest metallicity HII regions have upper limits $\mathrm{[N~II]}/\Ha < 0.04$,
so these are quite modest corrections.

The correction for [N II] reduces the \Ha\ flux while that for extinction increases it. The extinction correction is larger, so the net
correction is upwards, but only by a small amount. Combining both effects, we have corrected \Ha\ luminosities
$L_c(\Ha) = 1.15 L(\Ha)$ (lower Z) and $L_c(\Ha) = 1.26 L(\Ha)$ (higher Z). Rather than pretend that the second digit is significant,
we split the difference and apply a net upward correction of a factor of 1.2 to the entire sample. {In effect, we add 0.08 dex to eq.\ \ref{eq:KEsfr}.}

Our 20\% correction for extinction and [N II] contamination is small compared to the same correction in bright spirals where the HII
regions are both dustier and have greater [N II] emission \citep{McCall85,ZKH94}. 
It is also smaller than the systematic uncertainty in the calibration \citep{KE12,McQUVSFR}.
Overall, the low extinction and weak [N II] lines in LSB galaxies make for a relatively clean use of the \Ha\ line as a measure of the star formation rate.
Star formation rates computed in this fashion are given in Table \ref{sfrdata}.

For the SINGS comparison sample \citep{K09} we update the SFR to the calibration of \citet{KE12}.
We correct for the [N II] flux reported for each individual object. To account for extinction, we use the 24$\micron$ flux \citep{Dale07} and 
the corresponding correction from Table 2 of \citet{KE12}.

\section{Relations Between Star Formation, Stellar Mass, and Gas Mass}
\label{sec:fits}

We fit relations between the masses and star formation rates of LSB galaxies to estimate the
run of properties with one another as well as the intrinsic scatter in these relations. Fit results are given in Table \ref{fitresults}.
Measurement errors are generally small \citep[see][]{SMI}, and certainly smaller than the systematic uncertainty 
in the calibration of stellar mass and SFR (see \S \ref{sec:systematics}).

\begin{table}[h]
\begin{deluxetable}{cccccc}
\tablewidth{0pt}
\tablecaption{Fitted Relations for LSB Galaxies\label{fitresults}}
\tablehead{
\colhead{$y$}  & \colhead{$x$}  & \colhead{$a$} & \colhead{$b$} & \colhead{$\sigma$} 
}
\startdata
\Mg & \Mst & $\phn \phn 2.16\pm0.47$ & $\phn 0.79\pm0.06$ & 0.31 \\[-2pt]
SFR\tablenotemark{a} & \Mst 	& $-10.75\pm0.53$ & $\phn 1.04\pm0.06$ & 0.34 \\[-2pt]
SFR\tablenotemark{b} & \Mg	& $-14.93\pm1.10$ & $\phn 1.47\pm0.11$ & 0.47 \\[-2pt]
SFR & \Mb	& $-12.43\pm0.65$ & $\phn 1.16\pm0.07$ & 0.42 \\[-2pt]
\enddata
\tablenotetext{{\rm a}}{The star forming main sequence.}
\tablenotetext{{\rm b}}{The global analog of the Kennicutt-Schmidt relation.}
\tablecomments{Fits of the form $\log(y) = a+b \log(x) \pm \sigma$ where $\sigma$ is the intrinsic scatter
in the relation. Mass is measured in \Msun\ and star formation rate in \sfrate.}
\end{deluxetable}
\end{table}

As expected, larger galaxies have more of everything.
Galaxies with higher baryonic mass have larger stellar mass, gas mass, and star formation rate.
Our sample is not complete, so one should not take the fitted relations as absolute.
Nevertheless, as the sample extends to lower surface brightness and lower mass than commonly explored,
the slope and scatter of the relations provide some insight into galaxy evolution.

\subsection{Gas Mass and Stellar Mass}

The relation between stellar and gas mass is shown in Fig.\ \ref{SFMS}.
The majority of the LSB galaxy sample is gas rich, with $\Mg > \Mst$. 
In contrast, the bright spirals of SINGS are relatively gas poor, with $\Mg < \Mst$.
They fall below the upwards extrapolation of the fit to the LSB data.

Indeed, there appears to be a change in behavior at the scale $\Mst \approx 10^{10}\;\Msun$.
Below this mass, LSB dwarfs form a sequence of gas rich galaxies.
At a first approximation, this sequence is not far different from a constant $\Mg \approx 2 \Mst$.
In detail, the fitted relation gradually approaches star-gas equality as mass increases.
Around $\Mst \approx 10^{10}\;\Msun$ the slope of the relation between stellar and gas mass
flattens and the scatter increases as we enter the regime of giant spirals. 

As a check whether this transition is an artifact of the meeting point of different samples, 
we employ the SPARC database \citep{SPARC}. SPARC does not have measured star formation rates, 
but it does have robust stellar and gas mass estimates for a large sample of disk galaxies spanning
a broad range of mass and morphological types from S0 to dIrr. 
SPARC galaxies fill the same space in the \Mg-\Mst\ plane
as do the galaxies with measured SFR as does the large sample of \citet{BGB2015}. 

There appear to be two sequences of galaxies in Fig.\ \ref{SFMS}:
gas rich dwarfs at low mass and star rich spirals at high mass.
The point of separation is $\Mst \approx 10^{10}\;\Msun$.
The two groups are also well separated by morphological type:
early type spirals ($T < 5$: Sbc and earlier) are exclusively the more massive galaxies  
while the late types ($T > 6$: Sd and later) are exclusively less massive dwarfs.
Intermediate types (Sc and Scd) appear on either side of the break,
though Sc galaxies are more frequently giants and Scd galaxies are more frequently dwarfs.

Late type dwarfs form a sequence of star forming galaxies with an ample supply of cold gas.
In contrast, the early type spirals have low gas fractions and may be near to exhausting 
their reservoir of gas available for star formation.
To extend the analogy of the star forming ``main sequence'' to Fig.\ \ref{SFMS},
it appears that giant galaxies are ``turning off.'' It is easy to imagine that
as their gas supply is exhausted, galaxies follow an evolutionary trajectory
of declining \Mg\ as they approach a maximum \Mst. Perhaps the characteristic
timescale for this evolution is a function of mass such that $\Mst \approx 10^{10}\;\Msun$
represents the current turn-off mass. Below this mass is a sequence of thriving dwarfs
with ample gas to continue forming stars far into the future, while above it we have the
weary giants whose star forming potential is mostly in the past.

\subsection{The Star Forming Main Sequence}
\label{sec:SFMS}

The relation between the current star formation rate as traced by \Ha\ and stellar mass is shown in Fig.\ \ref{SFMS}.
This diagram that has come to be called the ``main sequence'' of galaxies.
The LSB galaxies in our sample extend over four decades in star formation rate: $4 \times 10^{-5} < \SFR < 0.3 \;\sfrate$.

The large dynamic range of the LSB data provides a strong constraint on the much-debated slope of the main sequence.
The fit we obtain has a slope consistent with unity: $1.04 \pm 0.06$ (Table \ref{fitresults}).
This is also found in other works that extend over a large mass range \cite[e.g.,][]{Peng2010,Kurczynski}.
The very low mass galaxy Leo P nicely follows the extrapolation of our fit: 
[$\log(\Mst)$, $\log(\SFR)$] = (5.75, $-4.53$) \citep{LeoPunquenched}.
The \Ha\ fluxes of the lowest mass galaxies can be generated by a single O star, 
which manifestly exist at these low star formation rates \citep[cf.][]{WKPAV2013,IGIMF}. There is no
dearth of massive stars in LSB galaxies resolved with HST \citep[][]{SMV}. 

There are samples with much larger numbers of galaxies than we have here.
However, these are usually subject to a strong selection bias against low luminosity and low surface brightness
galaxies. Most works that measure slopes for the main sequence less than unity are dominated by galaxies
with high mass. These large samples are dominated by giants with $\Mst \gtrsim 10^{10}\;\Msun$, 
and usually truncate around $\Mst \approx 10^{9}\;\Msun$. Consequently, they are mostly sampling the regime of weary giants.

It is obvious by inspection of Fig.\ \ref{SFMS} that the weary giants from SINGS have a shallow slope. They all fall below
the extrapolation of the fit to the LSB galaxies. So our result does not contradict but rather compliments other work.
Indeed, the star forming main sequence has a similar morphology to that in gas vs.\ stellar mass:
the thriving dwarfs form the main sequence while the weary giants are turning off it, having star formation rates
that are lower now than in the past.

The axes of the star forming main sequence are measured independently, but they are not independent physical quantities.
The stellar mass is the integral of the star formation rate: 
\begin{equation}
\Mst = \int_{t_F}^{t_U} \SFR \, dt = \aveSFR  t_G
\label{eq:sfhist}
\end{equation}
where $t_F$ is the time of galaxy formation, $t_U$ is the age of the universe, $t_G = t_U-t_F$ is the age of the galaxy, and 
$\aveSFR  = \Mst/t_G$ is the average star formation rate.

The relation between SFR and its integral, the stellar mass, places a physical upper envelope on where galaxies can reside
in the \SFR-\Mst\ plane. The current SFR may fluctuate, but if $\SFR > \aveSFR$ for long, it simply builds up \Mst\ 
and increases \aveSFR. The upper envelope is thus defined by the line of constant star formation
\begin{equation}
\SFR \lesssim \Mst/t_G.
\label{eq:env}
\end{equation}
For $t_G = 13$ Gyr, this corresponds to the line
\begin{equation}
\log(\SFR) = \log(\Mst) -10.11.
\label{eq:line}
\end{equation}
No galaxy may persist for long above this line. Such a high rate of star formation, if sustained,
makes enough stars to drive the position of the galaxy back to this line.
There can of course be temporary excursions with $\SFR > \aveSFR$,
but these starbursts must be brief in proportion to their intensity.

Despite this hard physical limit, there are many measurements of the slope of the star forming main sequence that are less than one.
These lead to absurd results when applied to low mass galaxies.
To give a specific example, \citet{Speagle2014} obtain a slope that is a function of both mass and time. By the time the universe reaches
it present age, appropriate to our low redshift galaxies, they give $\log(\SFR) = 0.5 \log(\Mst)-5.08$.
This is plausible for massive galaxies, but not for low mass galaxies.
For the lowest mass LSB galaxies, $\Mst \approx 10^7\;\Msun$, this formula predicts $\SFR = 0.026\;\sfrate$. This lies in a region
of Fig.\ \ref{SFMS} that is completely devoid of data. It is over two orders of magnitude higher than the observed SFR for such low mass galaxies
($\lesssim 10^{-4}\;\sfrate$). A $10^7\;\Msun$ galaxy forming stars at $0.026\;\sfrate$ would form its entire stellar mass in
just 0.4 Gyr.  The only way for the slope of the star forming main sequence to be shallow for low mass galaxies is for them to form 
quite recently. This is not consistent with resolved color magnitude diagrams \citep{McQstarbursts,McQstarburstsII,Weisz14,SMV}.

The shallow slope reported by many studies is simply unphysical when extrapolated to low mass galaxies. 
Nevertheless, it is indeed a good description of the data to which it is fit. 
The difference appears to be attributable to a selection effect: 
it is the consequence of samples dominated by luminous, high stellar mass galaxies.
One can see this effect in the SINGS data in Fig.\ \ref{SFMS}: a line drawn through the most massive galaxies
would lead to a similar result when extrapolated to very low mass. 

It seems more appropriate to identify the main sequence as that defined by star forming dwarfs with unity slope. 
Massive galaxies appear to be turning off the main sequence,
evolving to lower SFR than the mean SFR that made them what they are \citep[see also][]{Oemler17}.
It is this quenching process that the time-varying slope of the meta-analysis of \citet{Speagle2014} quantifies,
not the main sequence itself, which must, per force of eq.\ \ref{eq:sfhist}, have slope one \citep{Kelson}.
Indeed, even among massive galaxies the quenching process may represent the build-up of bulge components,
with star forming disks remaining closer to the main sequence \citep{Abramson14}.

Quantitatively, the sequence defined by LSB dwarfs is slightly below the locus  
of constant SFR. The stellar mass is a bit greater than what would be made
in a Hubble time at the current star formation rate, but not by much. On average, the current SFR is about half
the past average. This is probably within the systematic uncertainty of the calibration of both stellar mass and SFR,
so to a first approximation, LSB galaxies are consistent with having a constant star formation rate over the age
of the universe. Taking the calibrations literally, they have slowly declining SFR, consistent with an exponential
decline with e-folding time $\tau$ comparable to a Hubble time. Indeed, the vast majority of the LSB data are
consistent with $11 < \tau < 13$ Gyr.
in the SFR on the short ($\lesssim 10^7$ year) timescales of O star lifetimes \citep[see Fig. 13 of][]{SMII}

\subsection{The Kennicutt-Schmidt Relation}

Just as the star formation rate is related to the mass of stars it forms, so too is it related to the mass of cold gas from which stars can form.
The relation between SFR and \Mg\ is shown in the lower left panel of Fig.\ \ref{SFMS}. 
This is a global version of the Kennicutt-Schmidt relation.
The Kennicutt-Schmidt relation is usually posed as a relation 
between the surface density of star formation and the surface density of gas, but these are identical
to Fig.\ \ref{SFMS} when integrated over the same area \citep{K89,K98}.

The relation between the star formation rate and gas mass for LSB galaxies in our sample is given in Table \ref{fitresults}.
The slope ($1.47 \pm 0.11$) is consistent with that found in previous studies \citep[e.g., 1.4 -- 1.5:][]{K89,K98,KE12}.
Unlike the local surface density version of this relation \citep[e.g.,][]{leroy}, there is no hint of a turn down at low mass 
as there is at low surface brightness --- even among very low surface brightness galaxies \citep[see also][]{FiggsKS}.
This is because we simply consider the total gas mass here, not the local surface density as in \citet{leroy}, combined with the fact that
there is a much greater dynamic range in stellar surface density than in HI surface density: low surface brightness in stars does not
mean proportionately low surface density in atomic gas.

The non-linear slope of the Kennicutt-Schmidt relation contrasts with the linear slope of the star forming main sequence for LSB galaxies.
So too does the scatter, which is greater in the global Kennicutt-Schmidt. {In this sense, the rate at which stars form is more closely related
to the mass of stars already present than to the mass of cold gas from which stars could form, at least for LSB galaxies.
This may reflect a distinction between active ($\mathrm{H}_2$) and inactive (HI) gas reservoirs. The bottleneck to star formation appears to be
the conversion of atomic to molecular gas \citep{GdB99}, not the overall supply of cold gas.}

{Looking now at the giant galaxies from SINGS as well as the LSB dwarfs, we see a continuous relation in the lower left panel 
of Fig.\ \ref{SFMS}. There is no apparent break in the global Kennicutt-Schmidt relation as we cross from the regime of dwarfs to giants.
This stands in contrast to the star forming main sequence, where giant galaxies fall systematically below the upwards extrapolation of the
fit to LSB galaxies. In this regard, gas mass is a more continuous and direct indicator of star formation rate than stellar mass,
despite the smaller scatter in the star forming main sequence of LSB dwarfs.}

{Many galaxy formation simulations utilize the Kennicutt-Schmidt to prescribe the rate of star formation.
These results provide a check on such prescriptions. The overall morphology of Fig.\ \ref{SFMS} should be reproduced.
In addition, it is important to check that such prescriptions have the correct amount of scatter in the global Kennicutt-Schmidt relation
without overproducing the scatter in the star forming main sequence.}

\subsection{Star Formation and Baryonic Mass}

For completeness, Fig.\ \ref{SFMS} shows the relation between the star formation rate and the baryonic mass ($\Mb = \Mst + \Mg$).
As expected, the slope and scatter in the \SFR-\Mb\ relation are intermediate between those of the \SFR-\Mst\ and \SFR-\Mg\ relations.
There is only a hint of the turndown in SFR at high mass that is apparent in the star forming main sequence.
{The combination of continuity and low scatter may make baryonic mass a good star formation rate
indicator for late type galaxies.}

\subsection{Systematic Effects}
\label{sec:systematics}

{Uncertainties in flux measurements are small \citep{SMI}, but systematic errors may not be. 
The conversion from 21 cm flux to gas mass is widely agreed, but different choices could be made for the 
conversion of $V$-band and \Ha\ luminosity to stellar mass and star formation rate. 
Here we illustrate the potential amplitude of such systematic uncertainties by checking different calibrations.} 

{Plausible changes to the stellar mass calibration seem unlikely to affect our results.
For example, the star forming main sequence we find for our adopted calibration has a slope 
$b = 1.04 \pm 0.06$ and intercept $a = -10.75 \pm 0.53$ (Table \ref{fitresults}).
If instead we adopt the \ML-color relation of \citet{IP2013} in place of eq.\ \ref{eq:portinari},
the slope and intercept of the star forming main sequence change to $b = 1.01 \pm 0.07$ and $a = -10.40 \pm 0.54$,
a change of $< 1 \sigma$. The basic result remains unchanged: 
a star forming main sequence of slope unity for thriving dwarfs, consistent with a roughly constant SFR over a Hubble time. 
Substantially larger changes in the stellar mass estimates are not plausible,
as they would violate both photometric \citep{MS14} and kinematic \citep{MS15,OneLaw} constraints.}

{By a similar token, there is some uncertainty in the conversion of \Ha\ luminosity to star formation rate \citep{KE12}.
The constant in eq.\ \ref{eq:KEsfr} may be metallicity dependent, as one expects
metal poor stars to be more efficient at producing UV photons than their higher metallicity counterparts.
This was considered, e.g., by \citet{Brinchmann2004}; from their Fig.\ 7 we estimate that the constant in eq.\ \ref{eq:KEsfr} 
may change by $\sim 0.07$ per dex in stellar mass as a consequence of the mass-metallicity relation.
This would change the slope of the star forming main sequence to $1.12 \pm 0.06$: slightly steeper,
but not significantly so.}

{It is also conceivable that there are systematic variations in the IMF \citep[e.g.,][]{Lee09,Meurer09} that change
the relation between \Ha\ luminosity and SFR. These may even
be expected on theoretical grounds \citep{PAWK07,PAK09}. However, we see no evidence for this in resolved color magnitude
diagrams \citep{SMV}, where it is possible to integrate the resolved stellar population, extrapolate with a normal IMF, and 
recover the correct total luminosity of LSB galaxies.} 

{We may nevertheless compute the effects of a systematic variation in the IMF. \citet{Lee09} suggest a correction to 
the \Ha\ SFR based on UV estimated star formation rates, which are themselves very uncertain \citep{McQUVSFR}. 
Updated to the calibration of eq.\ \ref{eq:KEsfr}, eq.\ 10 of \citet{Lee09} becomes
\begin{equation}
\log(\SFR) = 0.62 \{ \log[L(\Ha)]-41.27 \} -0.47.
\label{eq:LeeUV}
\end{equation}
This holds only for low \Ha\ luminosity galaxies with $\log[L(\Ha)] < 39.4\;\mathrm{erg}\,\mathrm{s}^{-1}$ ($\SFR < 0.014\;\sfrate$). 
Fully 36 of our 56 LSB galaxies fall below this threshold. Applying this formula
to our data leads to a star forming main sequence with a somewhat shallower slope: $b = 0.68 \pm 0.05$ and a correspondingly
different intercept $a = -7.48 \pm 0.38$. The slope found in this way is of course an artifact of eq.\ \ref{eq:LeeUV}.}

\begin{figure*}
\epsscale{1.0}
\plotone{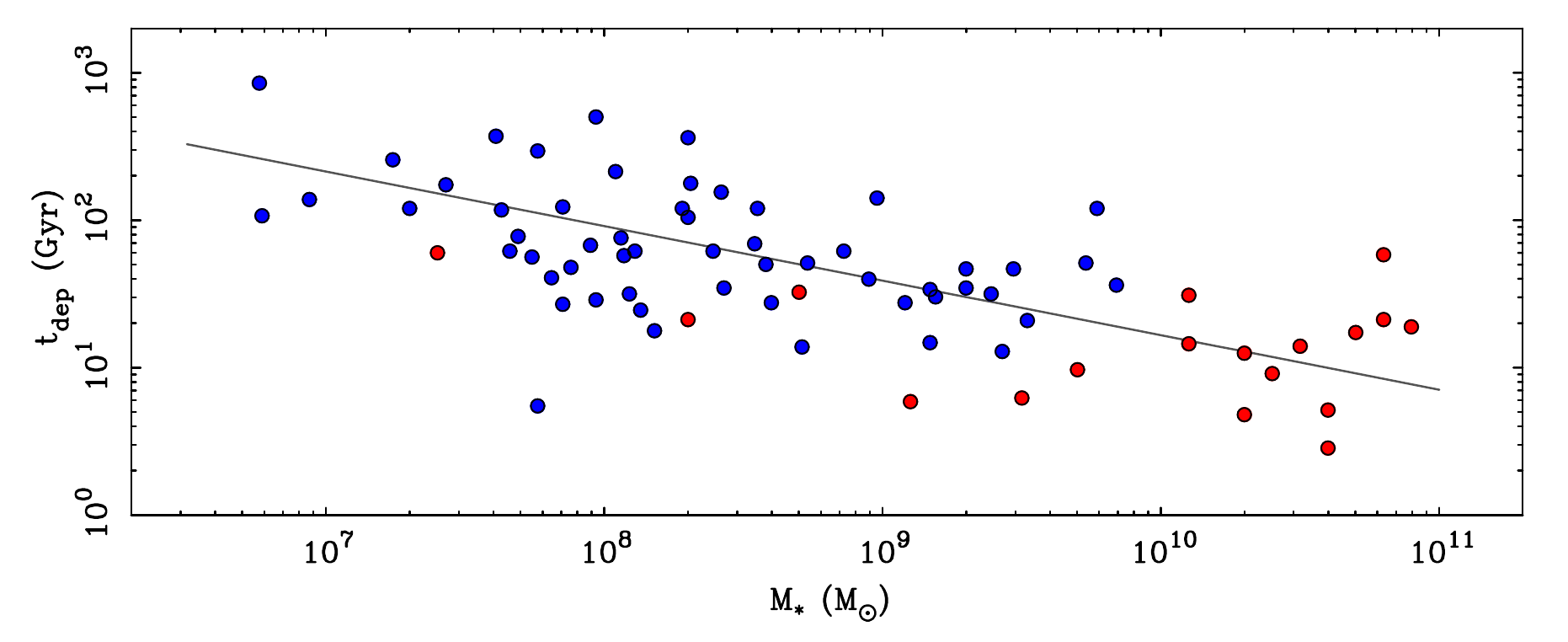}
\caption{The gas depletion time decreases steadily as a function of stellar mass.
The line $\log(t_{\mathrm{dep}}) = 4.92-0.37\log(\Mst)$ with $t_{\mathrm{dep}}$ 
in Gyr is derived from the relations for \SFR-\Mg\ and \Mg-\Mst\ in Table \ref{fitresults}.
Some of the weary giants have gas depletion times less than a Hubble time, but
thriving dwarfs may continue forming stars at their current rate for tens or even hundreds of Gyr without fresh accretion.
\label{tdepl}}
\end{figure*}

{The magnitude of the correction made by eq.\ \ref{eq:LeeUV} appears to be physically implausible, for the same reasons given
in \S \ref{sec:SFMS}: too many stars will form in less than a Hubble time. 
Prior to the application of this correction, most LSB galaxies have $SFR \lesssim \aveSFR$. Only 12 of the 56 have $SFR > \aveSFR$.
After the correction, 25 of the 36 affected galaxies have $SFR > \aveSFR$. Some are, by fiat, transformed into starbursts ($SFR \gg \aveSFR$). 
These LSB galaxies are manifestly not starbursts. }

{Our results appear to be robust against likely systematic effects.
Plausible systematic effects are comparable to the $\sim 1 \; \sigma$ uncertainties in the fits in Table \ref{fitresults}.
Large changes to these fits only occur for implausibly severe systematic effects.}

\section{Gas Depletion Times and Galaxy Evolution}

We can compute the time it will take to deplete the observed supply of cold gas at the current rate of star formation as $t_{dep} = \Mg/\SFR$.
This is shown in Fig.\ \ref{tdepl} as a function of stellar mass. The relation expected between these quantities can be derived from the fits
in Table \ref{fitresults}.

LSB galaxies have large HI reservoirs and low absolute star formation rates.
Consequently, their gas depletion times are long. These are typically tens to hundreds of Gyr.
There is no concern that these galaxies will exhaust their fuel for star formation anytime soon,
even by the cosmic standard of the Hubble time.

One consequence of the long depletion times of LSB galaxies is that there is no need to supply fresh material for star formation by 
external accretion. This may well happen, but there is no need to invoke it to maintain star formation. A brief depletion time is a
reason that is sometimes invoked for the need for external accretion in bright spirals, but these do not stand out in Fig.\ \ref{tdepl}. 
From the perspective of our study, the shorter depletion times of brighter galaxies may simply be an indication that their evolution
is nearer to completion in the sense that they have converted more of their original gas into stars. This is consistent with the
analogy to a main sequence turn off: star formation is slowing in massive galaxies simply as the supply of gas nears exhaustion.
There may be good reasons to believe galaxies continue to accrete substantial amounts of fresh gas, but short depletion times
is not one of them.

Rather than living in a special time, in which spirals are just now exhausting their gas supply, it may be that
weary giants with short depletion times merely mark the current phase of an on-going process.  
The red and dead early type galaxies of today may have been the
weary giants of the past, while the brightest dwarfs of the star forming main sequence may be the weary giants of the future.
As with the stellar main sequence, galaxies peel off in a mass dependent way, with the most massive galaxies completing
their star formation first.  This is, in effect, a rephrasing of the downsizing phenomenon, in which the most massive galaxies appear to
be the most evolved at any given redshift.  For the typical galaxy, quenching may simply follow from gas exhaustion.  The underlying
physical mechanism that makes this mass dependent is less clear.

\section{Conclusions}
\label{sec:conc}

We have examined the star formation rates in a sample of low surface brightness galaxies.
We find that dwarf LSB galaxies
\begin{itemize}
\item form a distinct sequence from massive spirals in the gas mass-stellar mass plane,
\item define a star forming main sequence of thriving dwarfs with unity slope,
\item are consistent with a global version of the Kennicutt-Schmidt law, and 
\item have gas depletion times much greater than a Hubble time, often tens of Hubble times.
\end{itemize}
There is no need to invoke fresh gas accretion to maintain star formation in LSB galaxies. It appears that
the \Mg-\Mst\ relation contains much of the same information as the \SFR-\Mst\ main sequence, albeit with greater scatter.
There is greater intrinsic scatter in the global Kennicutt-Schmidt law than in the star forming main sequence
{for thriving dwarfs}, which may have consequences for modeling galaxy evolution. 

More broadly, the star forming main sequence of galaxies may relate to its stellar namesake in a few qualitative ways.
Viewed from this perspective, we can identify three broad groups of galaxies:
thriving dwarfs (late morphological types of $\Mst < 10^{10}\;\Msun$), 
weary giants (early type spirals with $\Mst > 10^{10}\;\Msun$), 
and red and dead early type galaxies.
Thriving dwarfs are on the main sequence.  They have gas depletion times well in excess of a Hubble time, and can continue
to form stars at the observed rates far into the future.  Weary giants are still forming stars, and being massive galaxies, have the
highest absolute star formation rates among normal galaxies in the local universe.  However, their specific star formation rates 
appear to be in decline and they are in the process of turning off the main sequence. Red and dead early type galaxies have
practically ceased star formation and left the main sequence.  
This crude analog to stellar evolution can be seen in the unity slope of the star forming main
sequence for thriving dwarfs, the turn-down in slope for weary giants ($\Mst > 10^{10}\;\Msun$), and the offset of 
early type galaxies to very
low SFR at a given stellar mass.  This analog to the main sequence turn-off can also be seen in the \Mg-\Mst\ relation,
where the weary giants, in the slow process of leaving the main sequence, have gas masses lower than the extrapolation
of the relation for thriving dwarfs.  \\

\acknowledgements  We thank the referee for comments that helped us to clarify the presentation. This work is based in part on observations at Kitt Peak National Observatory, National Optical Astronomy Observatory which is operated by the Association of Universities for Research in Astronomy (AURA) under cooperative agreement with the National Science Foundation. The authors are honored to be permitted to conduct astronomical research on Iolkam Du'ag (Kitt Peak), a mountain with particular significance to the Tohono O'odham. 

\bibliography{SFMS}
\bibliographystyle{apj}

\end{document}